\documentclass{aa}
\usepackage{graphicx}
\usepackage{txfonts}
\usepackage{hyperref}
\usepackage[normalem]{ulem}
\hypersetup{colorlinks = true, citecolor = {blue}, urlcolor= {blue}}

\def\gapprox{\;\rlap{\lower 3.0pt                       
        \hbox{$\sim$}}\raise 2.5pt\hbox{$>$}\;}
\def\lapprox{\;\rlap{\lower 3.1pt                       
        \hbox{$\sim$}}\raise 2.7pt\hbox{$<$}\;}

\newcommand{\be}{ \begin{equation} }
\newcommand{\ee}{\end{equation}}

\newcommand{\ben}{\begin{enumerate}}
\newcommand{\een}{\end{enumerate}}

\renewenvironment{thebibliography}[1]{\begin{oldthebibliography}{#1}\setlength{\parskip}{0ex}\setlength{\itemsep}{0ex}}{\end{oldthebibliography}}

\usepackage{diagbox}
\usepackage{siunitx}
\newcommand{\orcid}[1]{\href{https://orcid.org/#1}{\protect\includegraphics[width=8pt]{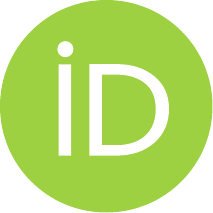}}}
\makeatletter
\renewcommand*\aa@pageof{, page \thepage{} of \pageref*{LastPage}}
\makeatother

\usepackage[dvipsnames]{xcolor}

\definecolor{darkgreen}{RGB}{31, 207, 31}

\usepackage{booktabs}
\newcommand{\N}{$N$}
\newcommand{\bonsai}{$\mathtt{bonsai2}$}
\newcommand{\phiGPU}{$\varphi\mathtt{-GPU}$}
\newcommand{\phiGRAPE}{$\varphi\mathtt{-\-GRAPE-\-hybrid}$}
\newcommand{\Msol}{\text{M}_\odot}

\begin{document}

    \title{Dynamics of supermassive black hole triples in the ROMULUS25 cosmological simulation}

    \author{
	    H. Koehn\inst{1,2}\orcid{0009-0001-5350-7468}
        \and
        A. Just\inst{1}\orcid{0000-0002-5144-9233}
	    \and
	    P. Berczik\inst{1,3,4,5}\orcid{0000-0003-4176-152X}
        \and
        M. Tremmel\inst{6}\orcid{0000-0002-4353-0306}
	}

\institute{
{Zentrum f\"ur Astronomie der Universit\"at Heidelberg, Astronomisches Rechen-Institut, M\"{o}nchhofstr. 12-14, 69120, Heidelberg, Germany, \email{\href{mailto:just@ari.uni-heidelberg.de}{just@ari.uni-heidelberg.de}}}
\and
{Institut f\"ur Physik und Astronomie, Universit\"at Potsdam, Haus 28, Karl-Liebknecht-Str. 24/25, 14476 Potsdam, Germany}
\and
{Nicolaus Copernicus Astronomical Centre, Polish Academy of Sciences, ul. Bartycka 18, 00-716 Warsaw, Poland}
\and
{Konkoly Observatory, Research Centre for Astronomy and Earth Sciences, E\"otv\"os Lor\'and Research Network (ELKH), MTA Centre of Excellence, Konkoly Thege Mikl\'os \'ut 15-17, 1121 Budapest, Hungary}
\and
{Main Astronomical Observatory, National Academy of Sciences of Ukraine, 27 Akademika Zabolotnoho St, 03143 Kyiv, Ukraine  \email{\href{mailto:berczik@mao.kiev.ua}{berczik@mao.kiev.ua}}}
\and
{School of Physics, University College Cork, Kane Building College Road, Cork, Ireland}
}

\date{Received 04 June 2023 / Accepted 16 August 2023}

\abstract{For a pair of supermassive black holes (SMBHs) in the remnant of a dual galaxy merger, well-known models exist to describe their dynamical evolution until the final coalescence accompanied by the emission of a low-frequency gravitational wave (GW) signal. In this article, we investigate the dynamical evolution of three SMBH triple systems recovered from the ROMULUS25 cosmological simulation to explore common dynamical evolution patterns and assess typical coalescence times. For this purpose, we construct initial conditions from the ROMULUS25 data and perform high-resolution gravitodynamical \N-body simulations. We track the orbital evolution from the galactic inspiral to the formation of hard binaries at sub-parsec separation and use the observed hardening rates to project the time of coalescence. In all cases, the two heaviest black holes form an efficiently hardening binary that merges within fractions of the Hubble time. The lightest SMBH either gets ejected, forms a stable hierarchical triple system with the heavier binary, forms a hardening binary with the previously merged binary's remnant, or remains on a wide galactic orbit. The coalescence times of the lighter black holes are thus significantly longer than for the heavier binary, as they experience lower dynamical friction and stellar hardening rates. We observe the formation of hierarchical triples when the density profile of the galactic nucleus is sufficiently steep.}


   \maketitle

\section{Introduction}
\label{sec:Introduction}
Massive galaxies experience a sequence of major mergers with other galaxies over the course of their lifetime. This simple fact, established by the $\Lambda$CDM model for hierarchical structure formation \citep{Peebles_1970, White_1978, White_1991}, has been confirmed through observations and numerical studies over the past decades \citep[see e.g.][]{Bell_2006, Naab_2006, van_Dokkum_2010, OLeary_2021}. Merging is an important process for galaxies to gain mass. At the same time, most elliptical and spiral galaxies contain a central supermassive black hole (SMBH) \citep{Kormendy_1995}. The empirical scaling relations between the masses of the SMBHs and the global properties of their host galaxies strongly suggest that SMBHs and their galaxies grow together. For instance, the  $M_{\text{BH}}$-$\sigma$ relation \citep{Silk_1998, Ferrarese_2000, AnnRev2013} links the velocity dispersion $\sigma$ of the central galactic bulge to the black hole mass $M_{\text{BH}}$. If the galaxy grows, $\sigma$ increases and so should subsequently $M_{\text{BH}}$. A similar logic applies to the $M_{\text{BH}}$-$M_{\text{Bulge}}$ or $M_{\text{BH}}$-$L_{\text{Bulge}}$ relations \citep{AnnRev2013, Graham_2013, Schutte_2019}, with $M_{\text{Bulge}}$ and $L_{\text{Bulge}}$ denoting the mass and luminosity of the central galactic bulge, respectively. Therefore, one assumes that during the course of a galaxy merger, the SMBHs that initially reside in the galactic nuclei of the progenitor galaxies, undergo a particular dynamical evolution that leads to the formation of an SMBH binary and its eventual coalescence under the emission of gravitational waves.

The basic sequence for this evolution was proposed by \citet{Begelman}, who suggested three different phases: First dynamical friction allows the black holes to sink down toward the center of the merger remnant, where they form a bound binary. Once the binary becomes hard, in the sense that its binding energy is larger than the average kinetic energy of the surrounding stars, dynamical friction is no longer effective but scattering encounters with stars crossing the galactic center extract energy from the SMBH binary and drive down their semi-major axis. For eccentric binaries, shrinking may also be caused by circumbinary disk accretion \citep{Franchini_2021, D’Orazio_2021, Siwek_2023, Lai_2023}, although for circular binaries the disk may even cause the separation to grow again in some settings \citep{Munoz_2019, Moody_2019}, but this remains matter of debate \citep{Heath_2020}. Eventually, the SMBHs come so close that gravitational wave (GW) emission leads to the final coalescence and emission of a sub-Hertz GW signal, which would be observable with the LISA satellite.

Various studies have verified this picture through analytic and numerical means \citep[see e.g.][]{Hills_1983, Ebisuzaki_1991, Quinlan_1996, Sesana_2006, Sesana_2008, Just_2011, Khan_2012, Khan13, Khan16, Bonetti_2020, Gualandris_2022}, for a review see \citet{Colpi_2014}. These efforts include the resolution of the infamous final parsec problem \citep{Milosavljevic_2003, Berczik05, Berczik_2006, Gualandris_2017, Kelley2017}. However, the timescale from the merger of the galaxies to the coalescence of the black holes can vary significantly depending on the remnant's matter distribution and the binary parameters, with timescales reaching from $50$ Myr up to $1$ Gyr \citep{Khan_2012, Colpi_2014, Khan16, Gualandris_2022}. The merger rate of massive galaxies in the local universe is on the order of $0.01$ Myr$^{-1}$ per Galaxy \citep{Mantha_2018, Duncan_2019} and increases up to redshift $z\sim 3$ \citep{Romano_2021}. Hence, it is very possible that a third galaxy joins a galaxy, which still contains an SMBH binary and creates a triple SMBH system, or that even three (or more) galaxies directly merge simultaneously. In fact, over the last years, about a dozen triple active galactic nuclei (AGN) systems have been observed \citep{Djorgovski_2007, Deane_2014, Liu_2019, Pfeifle_2019, Kollatschny_2020, Yadav_2021, Williams_2021}, with efforts for a systematic query of AGN databases for triple systems on their way \citep{Hou_2020, Foord_2021, Foord_2021_2}.

Therefore, the question arises of how such a supermassive black hole triplet evolves during and after a major galactic merger, as their dynamics could play out very differently compared to a dual SMBH merger. On the one side, one may assume that the SMBHs form a series of subsequent binaries in a stable hierarchical configuration and coalesce within the usual timespan \citep{Hoffman_2007, Ryu_2017}, whereas the other extreme possibility would be a chaotic evolution in which one or even all the black holes experience strong slingshots and may even get ejected from the center. Cosmological simulations are not capable of resolving SMBH dynamics down to parsec-scale, since their spatial and mass resolution is too low by orders of magnitude. 
For this reason, follow-up simulations with higher resolution are required to properly investigate the SMBH dynamics in galactic nuclei. While the question of how three-body dynamics affect the merger rates of stellar mass black holes has acclaimed increased attention \citep{Naoz_2013, Toonen_2016, Bhaskar_2022}, for SMBH triples, only a handful of few-body studies \citep[see e.g.][]{Valtonen_1989, Valtonen_1996, Blaes_2002, Hoffman_2007, Bonetti_2016, Ryu_2017, Bonetti_2018} and even less \N-body simulations \citep{Iwasawa_2006, Amaro_2010, Kulkarni_2012, branislavphd, Mannerkoski_2021} have so far been reported in the literature. Cosmological simulations have been the basis for the initial conditions only in \citet{branislavphd} and \citet{Mannerkoski_2021}. The former used IllustrisTNG100 as a starting point and resimulated six triple systems in \bonsai\ with higher resolution, but in none of the cases, a formation of a bound SMBH system was witnessed, as the SMBHs stalled at kiloparsec separation. In the latter article, the authors selected a region from their own large-volume, dark-matter-only simulation in which three halos collided and resimulated it with increased resolution in the smooth-particle-hydrodynamics $\mathtt{KETJU}$ code, to find that two black holes merged $\sim 3$ Gyr after the initial galaxy merger with the third black hole bound at parsec separations.

In this work, we present gravitodynamical \N-body simulations of three triple SMBH systems with initial conditions based on the ROMULUS25 cosmological simulation  \citep{Tremmel_2017}. By using the \bonsai, \phiGPU, and \phiGRAPE\ \N-body codes, we trace the orbital evolution of the SMBHs from kiloparsec down to sub-parsec separations. The article is organized as follows: The next section describes our search method for identifying triple systems in ROMULUS25 together with the creation of our initial conditions. Afterward, we briefly repeat some important aspects of SMBH orbital evolution in galaxies. Then we present the results of our simulations and provide estimates for the black hole merger timescales in each case before we continue with the discussion and conclusion of our study.
\section{Identifying SMBH triples in ROMULUS25 and creating initial conditions}
\label{sec:ROMULUS_Initial}
The ROMULUS simulations \citep{Tremmel_2017} are a suite of cosmological simulations with a particular emphasis on SMBH physics. Unlike in many other major cosmological simulations, SMBH particles are not artificially dragged to the potential minimum of their host halos but can move freely and are subject to a sub-grid dynamical friction force \citep{Tremmel_2015}. Additional advantages of the ROMULUS simulations include improved models for SMBH seeding, accretion, and feedback. For a detailed explanation, we refer to \citet{Tremmel_2017}. The SMBH masses in ROMULUS are proportional to the stellar mass of their host galaxies and follow the $M_{\text{BH}}$-$\sigma$ relation \citep{Tremmel_2017, Ricarte_2019}. The main run of the suite is ROMULUS25, containing a $25$ cMpc sized cube run to $z=0$ with a $0.35$ kpc spline kernel force softening equivalent to a Plummer softening of $\varepsilon_{\text{ROM}} = 0.25$ kpc and particle mass resolution $m_{\text{DM}} = 3.39\cdot 10^5\ \Msol$. Given this resolution, the dynamics are expected to be resolved up to scales $\sim \varepsilon_{\text{ROM}}$, and the proper SMBH orbital evolution can not be traced down until the merger. Instead, two SMBHs in ROMULUS are merged when their separation $\Delta r_{\text{BH}}$ is below $0.7$ kpc and their relative velocity $\Delta v_{\text{BH}}$ and acceleration $\Delta a_{\text{BH}}$ fulfill $\Delta v_{\text{BH}}^2 /2< \Delta a_{\text{BH}}\ \Delta r_{\text{BH}}$.

To properly resolve the SMBH dynamics after the galactic merger, we recreated ROMULUS25 systems with higher time, space, and mass resolution. For this purpose, we first performed a query of the ROMULUS25 SMBH merger tree to identify interesting triple systems. As a result, we retrieved three instances of such systems and then rebuilt the involved galaxies as spherical models, following a procedure similar to 
\citet{branislavphd}.

\subsection{Search criteria for SMBH triples in ROMULUS25}
\label{subsec:Search_criteria}
There are about 5500 SMBH merger events in ROMULUS25 in total. To identify which of these events could belong to triple systems, we checked whether the remnant SMBH of a given dual merger merges again with a third black hole during the next $1$ Gyr. We were interested in relatively isolated, high-mass SMBH systems, so we additionally applied the following criteria:
\begin{itemize}
    \item The masses of all SMBHs at their merger have to be larger than $10^7\ \Msol$.
    \item All mergers must occur after $z = 5.6$, to exclude spurious merger events when two seeds form close and merge quickly. 
    \item The two SMBHs merging initially must not have been involved in another merger for the past gigayear.
    \item The third black hole that later merges with the remnant from the first merger must not have been involved in another merger for the past gigayear.
    \item The remnant of the triple system must not engage in an additional merger for the coming gigayear.
\end{itemize}
After applying this selection method to the ROMULUS25 merger set, we found three triple SMBH systems, which we herein label systems A, B, and C. The ensuing simulations are referred to correspondingly. The properties of the two dual mergers constituting each triple event are recorded from the ROMULUS25 data in Table~\ref{tab:CandidateTable}. By further investigating the history of the corresponding host galaxies, one finds that all three triple systems have quite different characteristics. While in A three galaxies with central $10^7-10^8\ \Msol$ SMBHs collide simultaneously, in system B only two galaxies merge, one containing two equally sized $10^7\ \Msol$ SMBHs from an earlier galaxy merger event, and the other a central $5 \cdot 10^7\ \Msol$ black hole. In C, two smaller $10^7\ \Msol$ SMBHs orbit around the central $10^9\ \Msol$ within a single galaxy for several gigayears, until they are merged around $z = 0.35$. Therefore, our initial conditions for A, B, or C model three, two, or one galaxies respectively. Systems like C, where smaller SMBHs reside on a wide orbit with large sinking time scales after being stripped of their own stellar compound in a previous long-ago galactic merger, are common in ROMULUS25 and indicate the existence of a significant population of wandering SMBHs \citep{Tremmel_2018, Ricarte_2021a}.
\begin{table}
\caption{ROMULUS25 SMBH mergers constituting the triple systems A, B, C.}
\begin{tabular}{ p {0.08 \linewidth}  p {0.14 \linewidth}  p {0.17 \linewidth}  p {0.17 \linewidth} p {0.17 \linewidth}}
	\toprule
	   & Property & System A & System B & System C\\
	\midrule
    First &$t$, Gyr & 7.21 & 7.89 & 9.78\\
	merger  &&&&\\
	  &BH ID &  1981809013 &1982615610 & 1981810355 \\
	  &$m_1$, $\Msol$ & $6.282\cdot 10^{8}$ & $1.07 \cdot 10^{7}$ &  $3.17 \cdot 10^7$\\
	  &&&&\\
	  &BH ID & 1982228270 & 1986028215& 1982357852 \\
	  &$m_2$, $\Msol$ &   $3.59 \cdot 10^{7}$ & $1.63 \cdot 10^{7}$ & $1.05 \cdot 10^{7}$ \\
	  \midrule
   Second  &$t$, Gyr & 8.18 & 8.38 & 9.91\\
   merger  &&&&\\
	 &BH ID&  1981809013 & 1981818590& 1981810355\\
	 & $m_1$, $\Msol$ & $9.199 \cdot 10^{8}$ & $5.36\cdot 10^{7}$& $4.31\cdot 10^{7}$ \\
	  &&&&\\
	  &BH ID&  1981812636 & 1982615610& 1981814060\\
	  & $m_2$, $\Msol$ & $1.331 \cdot 10^8$ &$2.94 \cdot 10^{7}$ & $1.604\cdot 10^{9}$ \\
	\bottomrule
\end{tabular}
\label{tab:CandidateTable}
\end{table}

\subsection{Construction of the initial conditions}
\begin{table}
     \caption{Timespans and black hole masses in the simulations.}
     {\small
    \begin{tabular}{p{0.1 \textwidth} p{0.045 \textwidth} p{0.045 \textwidth} p{0.045 \textwidth} p{0.045 \textwidth} p{0.045 \textwidth}}
    \toprule
    Property & A.1 & A.2 & A.3 & B & C \\
    \midrule
    $t_{\text{start}}$, Gyr  & 6.25 & 6.25 & 6.25 & 7.77 & 9.49\\
    $t_{\text{end}}$, Gyr & 7.52 & 7.25 & 7.29 & 8.74 & 10.49\\
    \midrule
    $m_{\text{BH1}}$, $10^{7}\ \Msol$ & 88.4 & 88.4 & 88.4 & 5.4 & 160.4 \\
                     & (56.6) & (56.6) & (56.6) & (4.8) & (159.2)\\
    $m_{\text{BH2}}$, $10^{7}\ \Msol$ & 13.3 & 13.3 & 13.3 & 1.9 & 3.3\\
                     & (8.1) & (8.1) & (8.1) & (1.6) & (3.2)\\
    $m_{\text{BH3}}$, $10^{7}\ \Msol$ & 3.6 & 3.6 & 3.6 & 1.1 & 1.1\\
                     & (3.5) & (3.5) & (3.5) & (1.1) & (1.0)\\
    \bottomrule
    \end{tabular}}
    \tablefoot{The mass values in brackets correspond to the ROMULUS25 black hole masses at the initial time $t_{\text{start}}$, while the values actually adopted account for the accretion that happens in ROMULUS25 over the lifetime of the triple (see Sec.~\ref{subsec:Construction_of_IC}).}
    \label{tab:IC_parameters_global}
\end{table}

\begin{table}
    \caption{Galactic parameters for the initial conditions.}
    {\small
    \begin{tabular}{p{0.05 \textwidth} p{0.05 \textwidth}p{0.05 \textwidth} p{0.03 \textwidth}p{0.03 \textwidth}p{0.03 \textwidth}p{0.03 \textwidth}p{0.03 \textwidth}}
    \toprule
    Galaxy & Species & Property & A.1 & A.2 & A.3 & B & C\\
    \midrule
        1 & DM & $a$ & 33.33 & 33.3 & 33.3 & 39.8 & 74.6 \\
         & & $M$ & 216.9 & 216.9 & 216.9 & 109.0 & 892.9 \\
         \rule{0pt}{3ex}
         & Stars & $a$ & 0.7 & 2.2 & 1.3 & 0.4 & 4.4\\
         & & $\gamma$ & 0 & 1 & 1 & 1.4 & 1.9\\
         & & $M$ & 6.5 & 10.4 & 7.1 & 5.2 & 2.6\\
         \rule{0pt}{3ex}
         & Gas & $a$ & 49.1 & 49.1 & 49.1 & 40.0 & 149.7 \\
         & & $M$ & 34.3 & 34.3 & 34.3 & 6.2 & 198.7 \\
    \midrule
        2 & DM & $a$ & 18.8 & 18.8 & 18.8 & 5.5 & - \\
         & & $M$ & 64.9 & 64.9 & 64.9 & 10.4 & -\\
         \rule{0pt}{3ex}
         & Stars & $a$ & 1.2 & 3.8 & 2.2 & 2.31 & -\\
         & & $\gamma$ & 0 & 1 & 1 & 2 & -\\
         & & $M$ & 4.5 & 9.1 & 5.2 & 2.3 & -\\
         \rule{0pt}{3ex}
         & Gas & $a$ & 9.1 & 9.1 & 9.1 & 3.2 & -\\
         & & $M$ & 2.3 & 2.3 & 2.3 & 0.7 & -\\
    \midrule
        3 & DM & $a$ & 3.1 & 3.1 & 3.1 & - & -\\
         & & $M$ & 10.4 & 10.4 & 10.4 & - & - \\
         \rule{0pt}{3ex}
         & Stars & $a$ & 0.4 & 1.0 & 0.6 & - & -\\
         & & $\gamma$ & 0 & 1 & 1 & - & - \\
         & & $M$ & 3.0 & 3.5 & 3.1 & - & -\\
         \rule{0pt}{3ex}
         & Gas & $a$ & - & - & - & - & -\\
         & & $M$ & - & - & - & - & -\\
    \bottomrule
    \end{tabular}}
    \tablefoot{Hernquist and Dehnen parameters for the different matter components in the galaxies of our initial conditions. The scale radii $a$ are given in kiloparsec, the mass values $M$ in $10^{10}\ \Msol$}
    \label{tab:IC_parameters}
\end{table}
\label{subsec:Construction_of_IC}
\begin{figure*}
	\centering
	\includegraphics[width = 1 \linewidth]{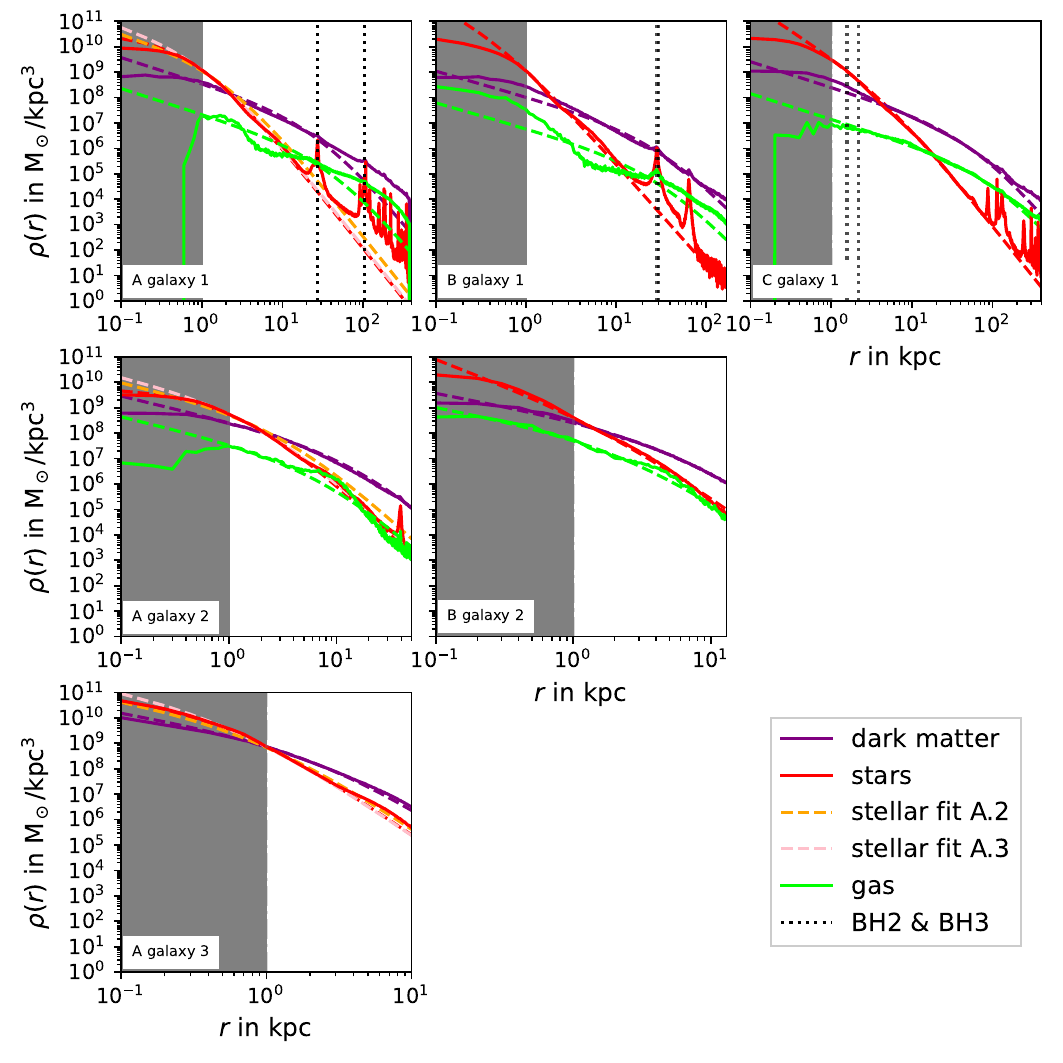}
	\caption{Radial density profiles of the galaxies from the three ROMULUS25 SMBH triple systems A (left column), B (middle column), C (right column). The solid lines show the actual ROMULUS25 data, while the dashed lines show fits with Hernquist and Dehnen models. Dark matter is color-coded purple, stars red, and gas green. The alternative stellar density fits for the initial conditions A.2 and A.3 are shown as dashed orange and pink lines respectively. Galaxy 3 in system A did not contain any gas. As stated, in system A the three SMBHs were hosted by three different galaxies, in system B by just two, and in system C by just one. The gray shaded areas were excluded from the fit, the dotted lines in the diagrams for the primary galaxies (top row) show the distance of BH2 and BH3 from their centers, where the heaviest black hole, i.e. BH1, of each triple system resides.}
	\label{fig:Initial_density_profiles}
\end{figure*}
\begin{figure}
	\centering
	\includegraphics[width = 1 \linewidth]{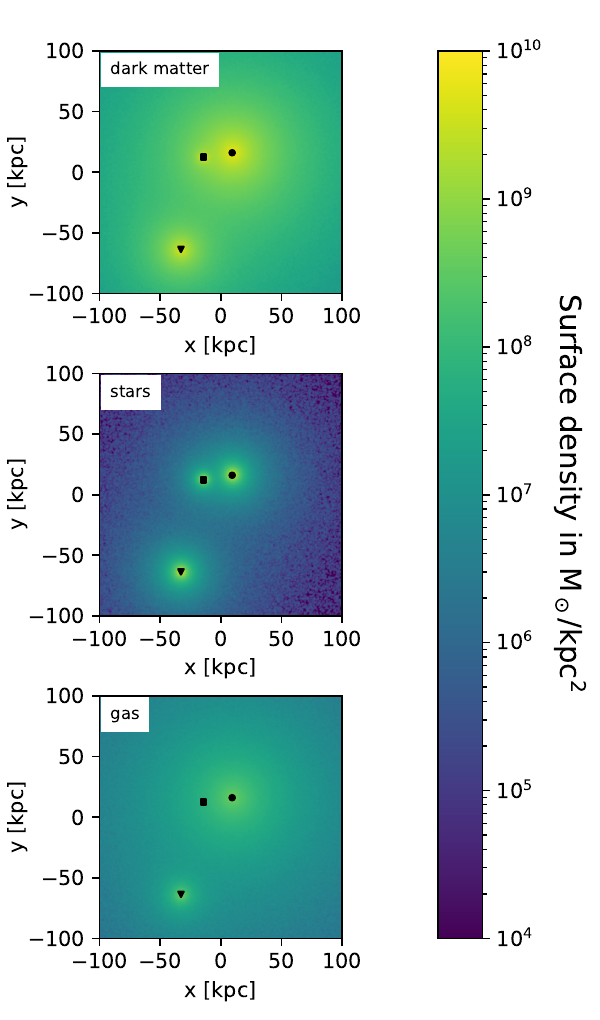}
	\caption{Surface density plot of initial condition A.1 reconstructed from ROMULUS25 system A. The top panel shows the distribution of the dark matter component, the middle panel the stars, and at the bottom the gas. The heaviest black hole BH1 is represented by the black dot, BH2 by the triangle, and BH3 by the square. Galaxy 3 did not contain any gas, therefore, no overdensity around BH3 is present in the bottom plot.}
	\label{fig:IC_Numa}
\end{figure}

To reconstruct the ROMULUS25 SMBH triple system, we had a ROMULUS25 database, constructed within the TANGOS framework \citep{Pontzen_2018}, at our disposal. This database contained radial density and mass profiles of the star, dark matter, and gas components for the galaxies of interest at various ROMULUS25 snapshots. Additionally, it also provided us with surface plots of the stellar and gaseous matter distribution for these respective galaxies. The galaxy models of our initial conditions are based on the radial density profiles and thus feature spherical symmetry. Naturally, this is a simplifying assumption, in fact, the surface plots from the database show galaxy 1 and 3 in system A, galaxy 1 in system B, and the only galaxy in system C as elliptical, while galaxy 2 in system A and galaxy 2 in system B display extended stellar disks around their central elliptical bulges. Likewise, the gas component actually forms disks in every case except in galaxy 3 of system A, where it is practically absent, and the one galaxy of system C. Hence, our initial conditions are to be seen as a first approximation to the real ROMULUS25 galaxies. We discuss the impact of the sphericity with respect to the outcomes in Sec.~\ref{sec:Discussion}. For systems A and B, we chose the starting point of our resimulation to be about 500 Myr before the first galactic merger in ROMULUS25, so the involved galaxies are still separated at that instance. In system C, the three SMBHs had already resided within the same galaxy for several gigayears as these SMBH mergers were not related to an immediate associated galactic collision. Therefore, we chose a starting point of 300 Myr prior to the first pair of black holes merging. The snapshots for the construction of our initial conditions were thus chosen at 6.25 Gyr, 7.77 Gyr, and 9.49 Gyr cosmic time for systems A, B, C respectively. Table~\ref{tab:IC_parameters_global} exhibits the start and end times of our resimulations.

We fitted the extended dark matter and gas profiles with the Hernquist model \citep{Hernquist_1990}, while the fit of the more compact stellar components relied on the Dehnen model~\citep{Dehnen_1993} 
\begin{align}
    \rho(r) &= (3-\gamma)\ \frac{M}{4 \pi}\ \frac{a}{r^\gamma (r+a)^{4-\gamma}}\text{,}\\
    m(r) &= \frac{M}{(a/r +1)^{3-\gamma}}\text{.}
\end{align}
These equations describe the density $\rho(r)$ and cumulative mass profiles $m(r)$ respectively.
The Hernquist model corresponds to a Dehnen model with $\gamma=1$. We excluded the inner $1$ kpc $= 4\varepsilon_{\text{ROM}}$ from the fit region, because there, due to the softening, the matter assembly and subsequently the densities were not reliable. 
In order to probe the effect of different extrapolations to the center, we constructed for system A three versions for the stellar density profile named A.1, A.2, A.3 (for details see subsection \ref{subsec:Central_density}).
The density profiles and resulting fits of all three triple systems are shown in Fig. \ref{fig:Initial_density_profiles}. Table \ref{tab:IC_parameters} shows the corresponding fit parameters. The inferred galactic parameters from the density fits were used in AGAMA \citep{Vasiliev_2019} to obtain 3-component \N-body models in dynamical equilibrium. The fits from the mass profiles were considered a secondary check and delivered consistent results. Initial positions and velocities of the SMBHs were taken from the ROMULUS25 orbital SMBH data. Fig. \ref{fig:IC_Numa} shows as an example the initial condition for A.1 as a surface density plot.

We encountered ambiguity for the SMBH mass values, because the black holes in ROMULUS accrete matter, but our dynamical \N-body codes do not account for SMBH growth. We decided to set the black hole masses to their values immediately before their last merger. In this way, we accounted in advance for the accretion that would occur in ROMULUS25 over the time of our simulation. Generally, in each simulation, the label BH1 is given to the heaviest black hole, BH2 to the intermediate one, and BH3 for the lightest one. Galaxies are labeled correspondingly.

\subsection{Density distribution in the galactic centers}
\label{subsec:Central_density}
Because the inner 1 kpc region of the galaxies had to be excluded from our fits, the proper density distribution of our initial conditions is poorly constrained in this region. Since this can have a tremendous impact on the eventual outcome, we created three distinct variations of system A, because having three separated galaxies, it appeared particularly susceptible to any changes in the inner galactic regions. These modified initial conditions and their simulations are herein called A.1, A.2, and A.3. In A.1 the stellar density was fitted without any prior constraints. It is shown in Fig. \ref{fig:IC_Numa} as the red dashed line. This resulted in a very flat stellar density profile with $\gamma = 0$. For A.3, $\gamma$ was fixed to 1 and $a$ and $M$ were fitted. For A.2, we again fixed $\gamma$ to 1 and additionally, the scale radius $a$ to the corresponding value from A.3 multiplied by $\sqrt{3}$, so the absolute densities in the central region in A.2 are roughly a factor of $3$ smaller than in A.3. Thus, one can characterize the different variations as A.1 having galaxies with very flat, low-density cores, while in A.3 the density profiles are cuspy and the central density is overall larger in magnitude. A.2 is an intermediate case, where the stellar distribution also has a steep cusp, but lower absolute central density. In all variations, the fits approximate the ROMULUS density at $r>1$ kpc reasonably well, as can be seen in Fig.~\ref{fig:Initial_density_profiles} (red, orange, and pink dashed lines in the first column). The three galaxies in A.2 have slightly higher stellar masses than in A.1 or A.3. For reference, the stellar profiles in B and C were fitted without any prior constraint and lead to cusps with $\gamma =1.4$, $\gamma=2$, and $\gamma = 1.9$ respectively, see Table \ref{tab:IC_parameters}.

\section{Simulation method}
\label{sec:Simulation_method}
The initial conditions for systems A and B were created with $N \approx 25\times 10^6$ particles, while for C we had $N \approx 15\times 10^6$. We began the simulations with the fast oct-tree code \bonsai, a custom version of $\mathtt{bonsai}$ \citep{Bedorf_2012b, Bedorf_2012}. As long as the SMBHs had large separations, in particular during the galactic inspiral phase in systems A and B, the uniform timestep symplectic integrator in \bonsai\ was able to track the dynamics accurately. Once the separation of one black hole pair fell below $\sim100$ pc, the risk of close pericenter passages not being properly resolved with a uniform timestep started to grow, so we resorted to other codes with an individual timestep scheme. Specifically, we relied on the gravitodynamical \N-body codes \phiGPU\ \citep{PHIGPU, Sobolenko2021, Sobolenko2022, Ber2022} and \phiGRAPE\ \citep{Meiron_2014, branislavphd}. \phiGPU\ is a direct \N-body code with a fourth-order hermitian integrator, where pairwise force calculation is accelerated by performing the computation on Graphic Processing Units (GPUs). Instead, \phiGRAPE\ emulates the $\mathtt{\varphi-GRAPE}$ program \citep{Harfst_2007}, originally written for special purpose GRAPE computers, on GPUs and additionally relies on the self-consistent field method \citep{Hernquist_1992} to compute the force for a large share of particles, though forces for specially selected "core" particles and the SMBHs are all calculated directly. Because of the approximation through the self-consistent field method, a significant amount of computation time, up to a factor of 16 compared to direct force calculation, can be saved and artificial relaxation in the outskirts is avoided \citep{branislavphd}. On the downside, \phiGRAPE\ is only well suited for centralized, roughly spherical systems. For the late phases, \phiGRAPE\ was used in A.2, B, and C, whereas the late phases in A.1 and A.3 were conducted solely with \phiGPU. This is because A.1 and A.3 were the first simulations to be performed, and we wanted to assess the capability of the newly adapted hybrid code by resimulating segments of A.1 and A.3 with it and subsequently compare the results. In A.2 and B \phiGPU\ was also used to bridge the times when the galaxies had not yet reassembled into a regular remnant, but the black hole separation was too small for the symplectic integrator in \bonsai\ to properly resolve the dynamical timescale. The sequence of code usage is also shown in Fig~\ref{fig:BH_separations}. As both \phiGPU\ and \phiGRAPE\ can only handle at most 6 million particles, we performed a radial cut around the center of the most massive galaxy, reducing the particle number to $4-6$ million upon switching to either code. \phiGRAPE\ in its current version does not support post-Newtonian (PN) terms in triple SMBH systems, hence runs A.2, B, and C are purely Newtonian. A.1. and A.3 relied only on \phiGPU\ at late phases, and there binary PN terms \citep{Sobolenko2017} up to order 2.5 were switched on when two SMBHs became bound. Force softening was set globally to $\varepsilon = 10^{-3}$ kpc in \bonsai, and upon code-switching reduced to $\varepsilon = 10^{-4}$ kpc for particle-particle interactions and $10^{-6}$ kpc for the SMBH interactions. A simulation was terminated, when a stable SMBH configuration had been reached, i.e. a hard binary or a stable hierarchical triple, and no further interesting evolution could be expected. The estimated merger time was then obtained by projecting the semi-major axis and eccentricity through Eqs.~(\ref{eq:GW_hardening}) and (\ref{eq:ecc_evol}), see Sec.~\ref{Sec:Theory}. In simulation B, we had to preemptively merge the particles representing BH1 and BH2 to prevent a computational halt by the decreasingly small orbital period of the binary. We projected the orbital elements of the preemptively merged binary until their expected physical merger, using again Eqs. (\ref{eq:GW_hardening}) and (\ref{eq:ecc_evol}), while the rest of the simulation continued.
\section{Dynamics of bound SMBH systems}
\label{Sec:Theory}
We consider two SMBHs with masses $m_{\text{BH1}}>m_{\text{BH2}}$ bound, when their relative energy
\begin{align}
\mathcal{E} = \frac{\mu}{2} \Delta \vec{v}^2 - \frac{G M \mu}{r}
\end{align}
is negative. Here, $\Delta \vec{v}$ is the relative velocity, $\mu$ the reduced mass of the binary, and $M$ its total mass. Therefore we can assign each bound pair of SMBHs a (time-dependent) Keplerian semi-major axis $a$ and eccentricity $e$ via the usual formulae. Two SMHBs normally become bound if their separation falls below the influence radius defined through the cumulative stellar mass $m_{\star}$ surrounding the SMBH binary:
\begin{align}
     m_{\star}(r_{\text{infl}}) = 2\ (m_{\text{BH1}} + m_{\text{BH2}})
\end{align}
If the semi-major axis becomes smaller than the hardening radius, determined through the velocity dispersion $\sigma$ of the surrounding matter
\begin{align}
    a_h = \frac{G m_{\text{BH2}}}{4 \sigma^2},
\end{align}
the binary is considered hard and dynamical friction turns ineffective in bringing the two black holes closer together \citep{Antonini_2011, Just_2011, Merritt_2013}. Instead, three-body scatterings between the binary and the lighter field stars extract energy at a constant hardening rate $s_{\text{SH}} = \frac{d}{dt}\frac{1}{a}$, under the crucial assumption that there is a continuous supply of closely interacting stars. The efficiency of this process in real systems can be assessed with the dimensionless hardening rate \citep{Quinlan_1996, Sesana_2006}
\begin{align}
    H = \frac{\sigma}{G \rho}\ \frac{d}{dt} \left(\frac{1}{a}\right),
\end{align}
where $\sigma$ is again the ambient velocity dispersion and $\rho$ the density.

After our simulations were terminated, we projected the merger times of the remaining hard SMBH pairs by calculating the time evolution of the semi-major axis $a$ and eccentricity $e$ through the coupled differential equations \citep{Peter_1963, Peters_1964}
\begin{align}
\frac{d}{dt}\left(\frac{1}{a}\right) &= s_{\text{SH}} + \frac{64}{5} \frac{G^3 \mu M^2}{c^5 a^5} \ \frac{1+ 73/24\ e^2 + 37/96\ e^4}{(1- e^2)^{\frac{7}{2}}},    \label{eq:GW_hardening} \\
\frac{de}{dt} &= -\frac{304}{15}\ \frac{G^3 \mu M^2}{c^5}\ a^{-4}\  \frac{e + 121/304 e^3}{(1-e^2)^{\frac{5}{2}}}.
\label{eq:ecc_evol}
\end{align}
Here, the linear stellar hardening rate $s_{\text{SH}}$ is obtainable from the preceding simulation. The second term in Eq. (\ref{eq:GW_hardening}) is the energy loss by GW emission. Eq. (\ref{eq:ecc_evol}) describes the circularization of the binary under the emission of gravitational waves and ignores any changes in eccentricity from the stellar hardening process, since the latter is expected to be negligible for a hard binary \citep{Quinlan_1996, Bonetti_2020, Gualandris_2022}.

It is expected, that a bound three-body system from a cosmological and thereby somewhat random initial condition either evolves chaotically or establishes a hierarchical triple, with one inner, closely bound binary and the tertiary body orbiting the inner binary's center of mass at a larger distance. The stability of such a hierarchical triple, where the bodies with masses $m_1$ and $m_2$ form the inner binary, can be assessed through the Mardling parameter \citep{Eggleton_1995, Mardling_2001, Toonen_2016, Hao_2022}
\begin{align}
    Y_m  = C (1-0.3\ \iota_{12-3}/\pi) \left(\biggl(1+\frac{m_{\text{3}}}{m_{\text{1}} + m_{\text{2}}}\biggr) \frac{1+e_{12-3}}{\sqrt{1-e_{12-3}}}\right)^{\frac{2}{5}},
\end{align}
where $C = 2.8$ is an empirical constant, $\iota_{12-3}$ the inclination of the tertiary orbit to the binary plane, and $e_{12-3}$ the eccentricity of the third body around the inner binary's center of mass. The triple is considered stable, if the ratio of the outer pericenter $r_{p, 12-3}$ and the inner semi-major axis $a_{12}$ is larger than the Mardling parameter, i.e. $r_{p,12-3}/a_{12}>Y_m$. We encountered a bound hierarchical triple only in simulations A.2 and A.3.

\section{Results}
\label{sec:Results}
\begin{figure*}
    \centering
    \includegraphics[width = 17 cm]{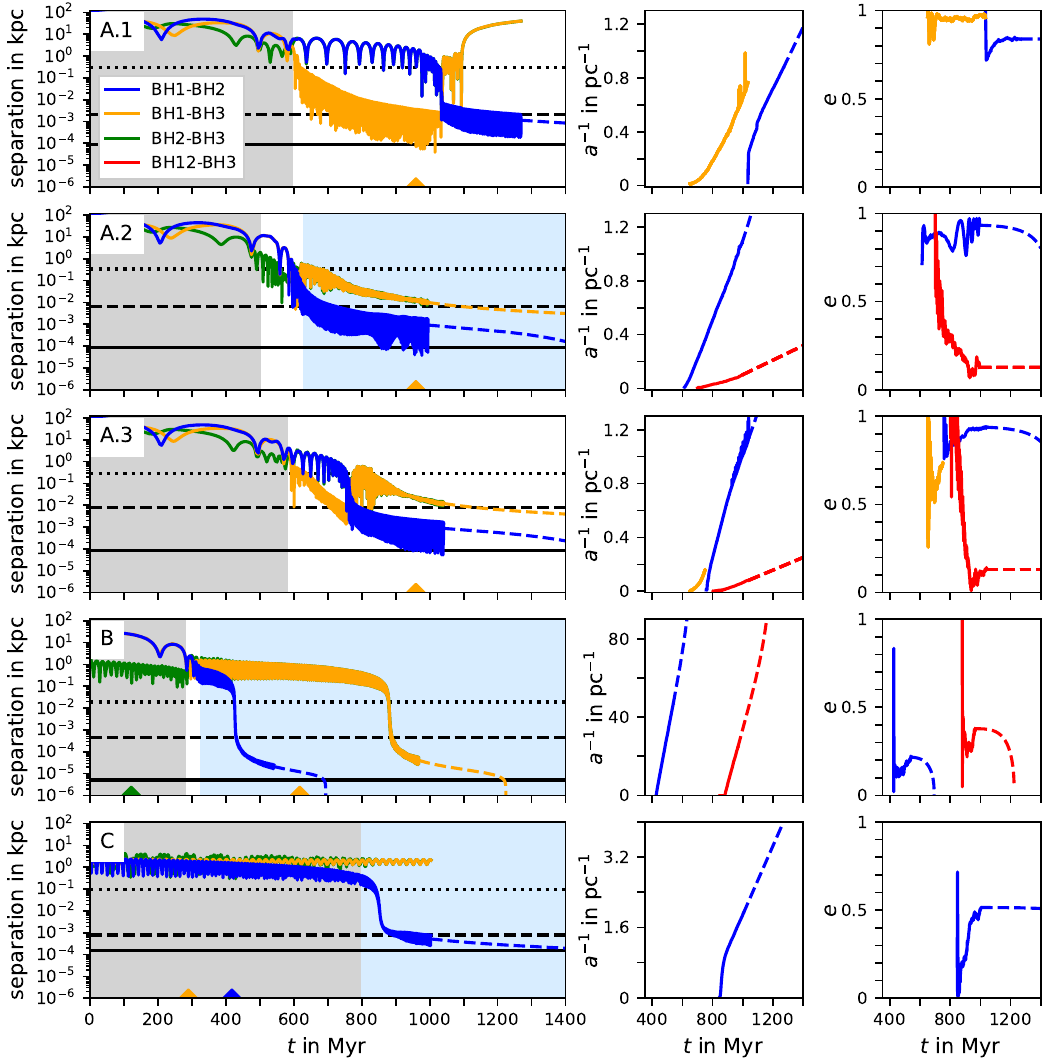}
    \caption{Black hole orbital parameters across the different simulations A.1 (top row) to C (bottom row). The color coding refers to different SMBH pairs. In A.2 and A.3, the red lines refer to the outer orbit of the hierarchical triple, in B the red lines refer to the BH3 orbit around the BH1-BH2 merger remnant. Background colors mark the codes used during the particular simulation period, light gray for \bonsai, white for \phiGPU, and light blue for \phiGRAPE. The left panels show the separation of different SMBH pairs. The black dotted, dashed, and solid lines respectively mark the influence radius, hardening radius, and the factor 1000 of the Schwarzschild radius of BH1. The color-coded bottom triangles mark the merger time point of an SMBH pair in ROMULUS25. The middle panels show the evolution of the inverse semi-major axis, the right panels the eccentricity evolution for bound SMBH systems. The colored dashed lines show the analytic projection of $a$, $a^{-1}$ and $e$ using Eqs. (\ref{eq:GW_hardening}) and (\ref{eq:ecc_evol}) after the simulation has been finished.}
    \label{fig:BH_separations}
\end{figure*}
While in A.1, A.2, A.3, and B the merger of the progenitor galaxies was simulated, C did not cover this phase because the black holes resided in the same ROMULUS25 halo for several gigayears. Hence, the nature of each system is quite different and we briefly characterize them separately in the following. The separations of the SMBH pairs in our simulations are shown over time in Fig. \ref{fig:BH_separations}, as well as the orbital parameter evolution for the bound binaries. 

In simulation A.1, with its initial three separated galaxies, galaxy 2 is tidally disrupted and partially merges with galaxy 3 after 400 Myr. BH2 and BH3 now share a common halo and orbit each other, but do not become directly bound. At 590 Myr, the remnant also collides and merges with galaxy 1, and a regular elliptical galaxy is formed. Here, BH2 experiences a slingshot by the incoming BH1 and is sent to an outer orbit with an apoapsis of 8 kpc and periapsis of 300 pc. At the same time, BH1 and BH3 sink to the center and form a bound, very eccentric binary that continues to harden. Meanwhile, the orbit of BH2 is subject to dynamical friction and at 1030 Myr it eventually descends into the influence sphere of BH1. There, a chaotic three-body interaction between all three SMBHs occurs, which results in the formation of a bound BH1-BH2 binary, while BH3, the lightest SMBH, is sent on a radial orbit and has repeated close encounters with the central binary, until at 1090 Myr, it experiences a very strong slingshot and is ejected from the stellar compound of the galaxy. Here, it should be noted that many of the outer particles had been removed after the final galaxy merger, so the potential well became shallower. Normally, BH3 would have remained bound to the dark matter halo, with a crossing time of $\sim 200$ Myr and weak dynamical friction. Even if another interaction between BH3 and the BH1-BH2 binary took place, from the mass ratios one would not expect another exchange but instead just another slingshot for BH3.

In simulations A.2 and A.3, representing the same ROMULUS25 system as A.1 but with different initial stellar density distributions, the galactic inspiral plays out very similarly. However, during simulation A.2, BH2 and BH3 approach each other more closely than in A.1, which then later during the final galactic collision weakens the BH2 slingshot. Here, BH2 immediately returns to the galactic center, preventing the formation of the BH1-BH3 system, and instead a BH1-BH2 binary forms, as BH3 is sent off to orbit around the galactic center at 100 - 500 pc. While the BH1-BH2 binary continues to harden, BH3 slowly descends, its orbit circularizes and a stable hierarchical triple is formed, where BH3 is bound to the inner BH1-BH2 binary. This is also the outcome in A.3, though BH2 needs a bit longer than in A.2 to fall back to the galactic center after the final galaxy merger, because here, due to lower stellar galactic masses, BH2 and BH3 do not approach each other as closely as they do in A.2. Hence, the slingshot at the final galaxy merger is stronger and for a short period of 100 Myr, BH2 is sent onto a 200 - 2000 pc orbit, while BH1 and BH3 actually form a bound binary. The latter is then disrupted by the descent of BH2 in a similar manner to A.1. However, in A.3 the ensuing chaotic three-body interaction results in the formation of a hierarchical triple. The two hierarchical triples we encountered at the end of our simulations A.2 and A.3 are stable with a Mardling parameter of $Y_m \approx 3$ and $r_{p,12-3}/a_{12} \approx 40$ in both cases.

In simulation B, two galaxies merge, where galaxy 1 hosts BH1 at its center and galaxy 2 contains BH2 and BH3 orbiting each other at distances of 100 - 1500 pc. The two galaxies collide and begin to merge after 290 Myr, which in turn separates the close (but unbound) BH2-BH3 pair. BH1 quickly sinks to the center of the remnant. BH2 and BH3 are on a wider galactic orbit, although BH2 is significantly closer with an apocenter of 800 pc compared to BH3's apocenter of 2000 pc. Hence, BH2 experiences more dynamical friction and starts to form a bound binary with BH1 at 430 Myr. This binary quickly hardens and merges after 694 Myr. Because of the small orbital period, the computational effort grew significantly and we preemptively merged BH1 and BH2 when their semi-major axis reached below $2 \cdot 10^{-2}$ pc to prevent a halt of the computation. Instead, the blue dashed curve in Fig. \ref{fig:BH_separations} shows the projected orbital parameters of BH1 and BH2 based on Eqs. (\ref{eq:GW_hardening}) and (\ref{eq:ecc_evol}). Over the lifetime of the BH1-BH2 binary, BH3 experiences little dynamical friction and so its apocenter only decreases slowly. After 880 Myr, BH3 falls into the center and becomes bound to the BH1-BH2 remnant. Together they form a quickly hardening binary, that will eventually merge.

Simulation C did not cover the merger of the progenitor galaxies, so in the beginning, all SMBHs already reside in a common halo. Here, BH1 sits at the center and the lighter black holes BH2 and BH3 are on a wider orbit, both orbits having an initial apocenter at 2300 pc. However, the pericenter of BH2 is significantly smaller with 500 pc, whereas BH3 never approaches the center closer than 1400 pc. Therefore, BH2 becomes subject to dynamical friction, the peri- and apocenter shrink and after 840 Myr it reaches the influence sphere of BH1, and the two form a bound binary. In contrast, BH3 stays on a stable orbit with very little variation in its apo- and pericenter across the entire simulation. The hardening of the BH1-BH2 binary  decelerates, once it becomes hard around 900 Myr, but then settles to a constant rate so the BH1-BH2 binary will eventually merge.

Hence, in each of our simulations, the black holes arrange themselves such that the most massive black hole is at the center of the final galaxy, while the lighter two black holes wander through its outer regions. Eventually, the two heaviest black holes form a bound binary. The fate of the lightest black hole varies drastically. It either gets ejected from the stellar compound (A.1), forms a hierarchical triple (A.2 and A.3), merges after the coalescence of the initial BH1-BH2 binary (B), or experiences virtually no dynamical friction and thus never becomes bound to another black hole (C). 

\subsection{Hardening rates and projected merger times}
\label{subsec:Project_merger}
\begin{table}[t!]
\caption{Hardening rates and merger times of SMBH binaries.}
\label{tab:mergertimes}
\begin{tabular}{>{\centering\arraybackslash} p {2.7 cm }  >{\centering\arraybackslash} p { 0.7 cm} >{\centering\arraybackslash} p {0.7 cm} >{\centering\arraybackslash} p {0.7 cm} >{\centering\arraybackslash} p {0.7 cm}  >{\centering\arraybackslash} p {0.7 cm} }
\toprule
Property& A.1 & A.2 & A.3 & B & C\\
\midrule
 $s_{12}$ & 2.30 & 2.92 & 3.20 & 400.66 & 7.76 \\
 $H_{12}$ & 7.8 & 11.3 & 6.5 & 4.5 & 3.4 \\
 $t_{c,12}$ in Myr & 2680 & 1489 &1450 & 694* & 2058 \\
 \midrule
 $s_{12-3}$  & - & 0.54 & 0.47 & 286.96 & - \\
 $H_{12-3}$ & - & 2.8 & 1.0 & 4.1 & - \\
 $t_{c,12-3}$ in Myr & - & 16466 & 18341 & 1224 & - \\
 \midrule
 final galaxy merger & 625 & 610 & 623 & 280 & -\\
\bottomrule
\end{tabular}
\tablefoot{*Merger occurred during simulation;
The stellar hardening rates $s$ are given in kpc$^{-1}$ Myr$^{-1}$, $H$ is dimensionless, and the projected coalescence time $t_c$ is given in Myr after the start of the simulation. The top block displays these quantities for the final BH1-BH2 binaries. The middle block displays these quantities for the final BH3 orbit, i.e. for the hierarchical triple in A.2 and A.3 and for the orbit of BH3 around the BH1-BH2 remnant in B. The bottom line displays the time of the final galaxy merger in Myr after the start of the simulation.}
\end{table}
At the end of our simulations, we projected the merger time of the remaining BH1-BH2 binaries using Eqs. (\ref{eq:GW_hardening}) and (\ref{eq:ecc_evol}). The results are displayed in Table \ref{tab:mergertimes}. In all simulations, the two heaviest black holes are projected to merge before any other pair. The time of coalescence after the final galaxy merger ranges from just 414 Myr in B to 2055 Myr in A.1, with dimensionless hardening rates between 3.4 to 11.3.

In A.2 and A.3 where BH3 eventually form a hierarchical triple with the BH1-BH2 binary, we also used Eqs. (\ref{eq:GW_hardening}) and (\ref{eq:ecc_evol}) for the extrapolation of its orbit around the BH1-BH2 binary's center of mass. Likewise, we used this method for the orbit of BH3 around the merged BH1-BH2 remnant. We observe that the inner BH1-BH2 binaries have higher absolute hardening rates than the BH3 orbits. In combination with the smaller mass ratios of the BH12-BH3 systems, this leads to significantly longer coalescence times for BH3, which in A.2 and A.3 even exceed the Hubble time. It should be noted that the estimates of $t_{c,12-3}$ in A.2 and A.3 are very conservative, in that they assume a constant, low hardening rate $s_{12-3}$ of the outer orbit, while in reality, the hardening might increase after the merger of the inner BH1-BH2 binary. This idea is supported by the fact that the dimensionless hardening rate $H_{12-3}$ is significantly smaller than $H_{12}$ in A.2 and A.3, but in B, where the inner binary has already merged, $H_{12}$ and $H_{12-3}$ are roughly equal. Therefore, it seems possible that the efficiency of the stellar hardening increases once the inner binary merges.
\subsection{Evolution of the matter distribution in the galactic center}
\begin{figure}
    \centering
    \includegraphics[width=\linewidth]{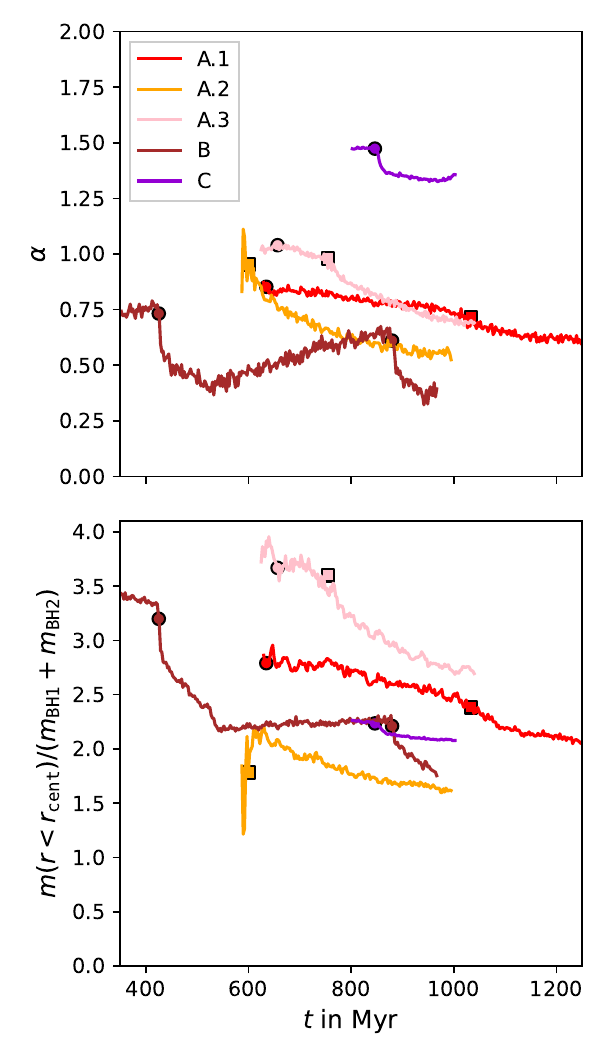}
    \caption{Time evolution of the central stellar density slope $\alpha$ (top) and total mass (bottom) inside the galactic center given in terms of the total binary mass. The slope and mass were determined in a sphere around the central binary. The sphere's radius $r_\mathrm{cent}$ was approximately the binary influence radius and set to 300 pc in A.1-A.3, 20 pc in B, and 100 pc in C. The different simulations are color-coded according to the legend. Sudden jumps in the slope and mass coincide with the beginning of direct chaotic triple SMBH interactions (squares) or the formation of bound binaries (circles).}
    \label{fig:matter}
\end{figure}
The interplay between the SMBHs and the galactic matter leaves an imprint on the shape of the galactic center. In Fig. \ref{fig:matter} we show the evolution of the power law slope $\rho \propto r^{-\alpha}$ of the stellar density profile and the total enclosed mass around the central binary. Clearly, the bound SMBH binaries remove matter inside their sphere of influence and flatten the density profile. This process is known as binary scouring and causes a mass deficit in inner galactic regions. The mass deficits here supersede the usual rule of thumb, whereby the removed mass should amount to half of the total binary mass \citep{Merritt_2006}, but are actually in agreement with merger remnants from other simulations \citep{Khan_2012, Dosopoulou_2021}.

The direct chaotic three-body interactions in A.1, A.2, and A.3 also trigger a sudden change in the slope and central mass. Here it is also apparent that the chaotic SMBH three-body interaction in A.1 takes place in a shallower density profile than in A.2 and A.3, whereas the absolute mass and density inside the influence radius are fairly different in all cases. Because these three simulations are based on the same ROMULUS25 triple system, but had variations in the central density distribution of the stellar component (see Sec.~\ref{subsec:Central_density}) the slope seems to exert a crucial influence on the outcome of the chaotic three-body interaction. We discuss this further below in Sec.~\ref{sec:Discussion}.

In B and C, no direct three-body interaction takes place, thus the depletion and flattening of the central density is only attributed to classical binary scouring. Noticeably in Fig.~\ref{fig:matter}, the flattening of the stellar density stops after BH1 and BH2 have been merged and is then even reversed as $\alpha$ starts to grow again until the second binary between the BH1-BH2 remnant and BH3 is formed. In contrast, the total amount of matter inside the center remains constant. The rebound of $\alpha$ is possibly linked to two-body relaxation after binary scouring has ended.

\subsection{Eccentricity evolution}
\label{subsec:eccentricity}
\begin{figure}
    \centering
    \includegraphics[width = \linewidth]{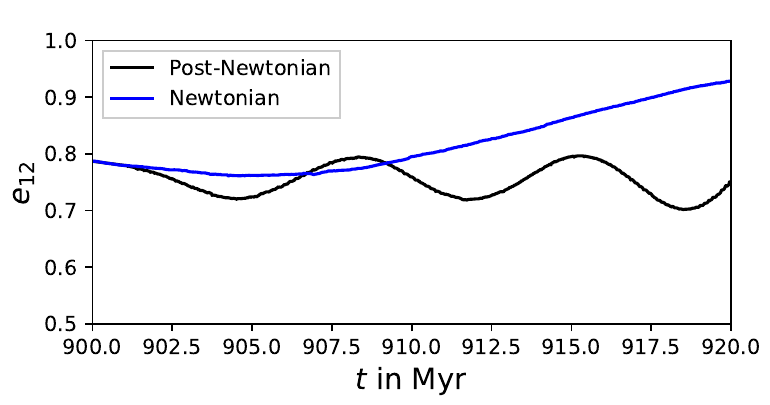}
    \caption{Eccentricity evolution of the inner binary in simulation A.2 for Newtonian gravity (blue) and when binary PN terms are switched on (black).}
    \label{fig:ecc_comp}
\end{figure}
The eccentricities usually grow only slightly after the formation of a given binary and then settle to a constant value. In A.1-A.3 the binaries have relatively high eccentricities above 0.8, while in B and C, the eccentricities do not supersede 0.5. According to \citet{Gualandris_2022} the binary eccentricity correlates with the orbital eccentricity around the galactic center before the two SMBHs become bound, so in a triple galaxy merger like in system A one may expect to see at times relatively high eccentricities. The outer orbit of the hierarchical triple in A.2 and A.3 quickly circularizes due to dynamical friction.

The Kozai-Lidov-von Zeipel (KLZ) effect has often been quoted to suggest that a third black hole in a hierarchical triple could excite the eccentricity of the inner binary and thus reduce its merger time \citep{Makino1994, Iwasawa_2006, Hoffman_2007, Naoz_2016, Gualandris_2017}. Indeed, we observe an oscillation of the inclination $\iota_{12-3}$ and some excitation of the inner eccentricity in simulation A.2 when the hierarchical triple is formed. We emphasize, that in the original KLZ mechanism, the secondary body in the inner binary is seen as a test particle, which is not applicable in our context. In any case, we can see that this excitation of the eccentricity does not occur if we switch on binary PN terms. Fig. \ref{fig:ecc_comp} compares the eccentricity evolution of the BH1-BH2 binary in simulation A.2, where only Newtonian gravity was present, to a brief subsidiary run, in which binary PN terms were switched on. We observe that the large fluctuation of the eccentricity is suppressed by the relativistic perihelion shifts, in agreement with previous studies \citep{Tanikawa_2011, Bonetti_2018, Mannerkoski_2021}. Similarly, we did not observe any KLZ-like effects in the hierarchical triple of simulation A.3, where PN terms were switched on for the inner binary because of the continued usage of \phiGPU.
\section{Discussion}
\label{sec:Discussion}
\begin{figure}
    \centering
    \includegraphics[width = \linewidth]{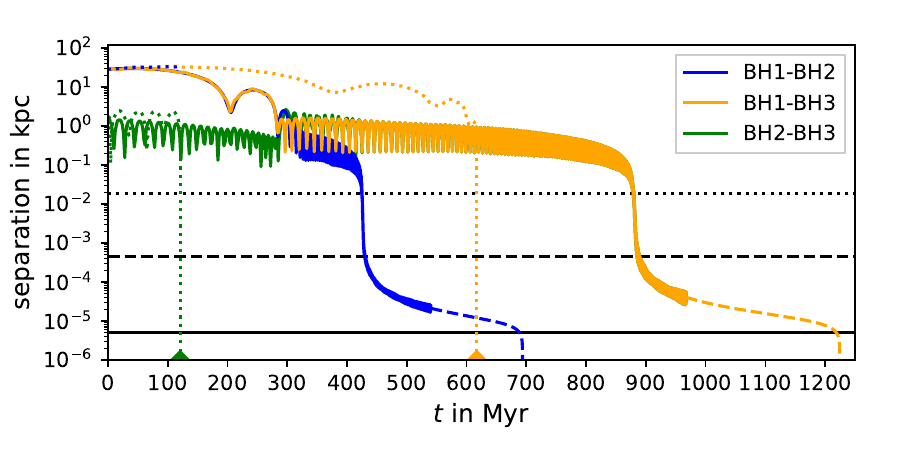}
    \caption{SMBH separation in simulation B as in Fig.~\ref{fig:BH_separations}. For comparison the SMBH separations in ROMULUS25 are shown as dotted lines. In ROMULUS25 the galactic inspiral takes about a factor of two longer than in our simulation and there BH2 and BH3 are merged before any interaction with the other galaxy took place.}
    \label{fig:ROMULUS_comp}
\end{figure}
The results of our simulations provide information on typical merger dynamics and timescales for SMBH triplets and their influence on the central density distribution. To assure the consistency of our results, we simulated the period from 630 - 840 Myr in A.1 and the period from 630 - 740 in A.3 twice, one time with \phiGPU\ as presented above, and the other with \phiGRAPE. We find very good agreement between both codes, in particular with respect to the binary hardening rates, but also in regards to the wider SMBH orbits and the evolution of the central density distribution. Specifically, the hardening rates are independent of the exact share of \phiGRAPE\ "core" particles (i.e. particles for which the self-consistent field method is \textit{not} applied), proving the ability of the hybrid code to accurately capture the star-SMBH interactions.

In our work, we model the ROMULUS25 galaxies as spheres and lack hydrodynamical treatment. The former is particularly relevant during the inspiral of the galaxies since the sphericity is destroyed over the course of the galactic merger because of the high galactic mass ratios. Deviations from sphericity in the original ROMULUS25 galaxies, in particular triaxiality, as well as extended stellar and gaseous disks enhance the susceptibility to tidal forces and influence the overall deformability, altering the orbital approach. When comparing our results to the original orbital evolution in ROMULUS25, one observes that in the cosmological simulation, the galactic inspiral takes about a factor of two longer. Fig.~\ref{fig:ROMULUS_comp} compares as an example the SMBH separations in simulation B to the separations in the original ROMULUS25 system. Such a premature galaxy merger compared against the ROMULUS25 merger timescale is typical in our simulations. However, despite the faster inspiral, we observe general congruence between our simulations and the dynamics in the cosmological simulation above the kiloparsec-scale. For instance, our simulations of system A replicate the BH2 slingshot at the final galaxy merger well, so in all cases BH1 and BH3 first form a close pair. Based on this observation, we also think that the impact of spherical approximations on the SMBH orbital evolution at the moment of the galactic merger is likely not very significant, although there the effects of triaxiality and disks have to be investigated further in future work. During simulation C a merger did not take place, the original ROMULUS25 galaxy is an older elliptical. Hardening rates may drop in spheric potentials, as fewer stars move on centrophilic orbits \citep{Berczik_2006, Khan13, Kelley2017, Gualandris_2017}, so when continuing simulation C for longer, one might have expected some decelerated hardening.

The lack of hydrodynamical phenomena in our work mainly impacts the stages that take place after the galactic mergers. Hydrodynamic effects might lead to the formation of a gaseous disk around the otherwise elliptical remnant. Inside the disk plane, SMBHs experience enhanced dynamical friction \citep{Bonetti_2020b, Szoelgyen_2021}, causing an earlier orbital plunge. However, \citet{Chen_2021} show how the gas drag itself only contributes very little to the actual dynamical friction. Neglecting hydrodynamics also means that no new stars are formed when gas falls into the center during a galactic merger. This influences the nuclear densities and thus the dynamical friction and hardening rates for the SMBHs, though recent studies indicate that the link between galaxy mergers and starbursts is not as strong as previously assumed and decreases with higher redshift \citep{Pearson_2019, Renaud_2022}. Zoom-in cosmological simulations of galaxy mergers and SMBH binary formation confirm that neglecting hydrodynamics when studying multiple SMBH systems still leads to realistic outcomes \citep{Mannerkoski_2021, Mannerkoski_2022}. Hence, we expect that our simulations cover the important aspects of SMBH triple evolution in dry galaxy mergers, in particular during the hardening phases, where we found that the order of the SMBH mergers changes when the interactions and exchanges neglected in the cosmological simulation are properly resolved.

In Sec.~\ref{subsec:Construction_of_IC}, we reported how we adopted the later ROMULUS25 SMBH mass values that account for accretion in advance since our codes can not model SMBH mass growth. Alternatively, one could have relied on the initial SMBH mass values from the ROMULUS25 snapshots at $t_{\text{start}}$. Both mass value sets are reported in Table~\ref{tab:IC_parameters_global}, the latter one in brackets. The largest accretion growths occur for BH1 and BH2 in system A with a mass increase of 40~\%, any other is smaller than 16~\%. The mutual mass ratios remain very similar. In early exploratory runs with \bonsai, we found no impact of the SMBH masses on the orbital evolution during the galactic inspiral, because at larger separations the SMBH orbits are determined through the surrounding stars. Only once the black holes got stripped of their stellar compounds, differences became apparent, but at that time a significant portion of the accretion would have already taken place. We did not test, how sticking with the lower mass values could have impacted the SMBH binding, but as we expect from the dynamical friction timescales the delay would be only on the order of 10-60 Myr. Likewise, we think stellar hardening rates would be mainly unaffected because the mass ratios remain unchanged \citep{Quinlan_1996, Sesana_2006}. The gravitational wave emission, however, would recede and in system A, the case with the highest accretion mass growth, the GW timescale would increase by a factor of 4 when adopting the lower SMBH masses. In the end, this would have inappropriately delayed the merger times in Table~\ref{tab:mergertimes} by 300-600 Myr for the binaries in A.1, A.2, or A.3.

From our results in Table \ref{tab:mergertimes} we see that the hardening rates fall short of the theoretical value $H \approx 15$ \citep{Sesana_2006, Bonetti_2020}. This is in agreement with studies on SMBH binary systems in realistic systems \citep{Sesana_2015, Lezhnin_2019}. We observe relatively quick coalescence of the heavy BH1-BH2 binaries within fractions of a Hubble time, which is also similar to previous studies on triple systems. There the heaviest black holes usually merge first and the lightest black hole, if at all, coalesces much later, exceeding the typical lifetime of 1 Gyr \citep{Iwasawa_2006, Hoffman_2007, Mannerkoski_2021}. There are several reasons for this general behavior. High-density, cuspy profiles accelerate the stellar hardening process \citep{Khan16, Khan_2018}. As a consequence, we have lower absolute hardening rates and lower stellar hardening efficiency for the outer orbits in our hierarchical triples when the galactic center has already been scoured by the heavier BH1-BH2 binary, as for instance in simulation B. However, as already mentioned in section \ref{subsec:Project_merger}, the hardening efficiency for the outer orbits in the triples of A.2 and A.3 could increase after the merger of the inner binary. Backing this claim would require further studies on the interactions between loss cone stars and the hierarchical SMBH triple, but is strongly hinted by the fact that in B the dimensionless hardening efficiencies for the BH1-BH2 binary and its remnant with BH3 are equal. At last, the longer coalescence times are also the result of the smaller mass ratios of BH3 to the BH1-BH2 remnants, which prolongs the gravitational wave emission phase.

A particularity in our results are the drastically different fates of BH3 in the variations of system A. Only in A.2 and A.3 a stable hierarchical triple forms after the three SMBH directly have chaotic interactions with each other, while in A1 it gets ejected from the stellar compound. The explanation for this strong difference is threefold. On the one side, in A.1 BH2 experiences a stronger slingshot during the final galaxy merger than in A.2 or A.3, because it got stripped earlier of its original, low-density stellar cusp. Secondly, in A.1 dynamical friction for BH2 is weaker, so the chaotic triple phase occurs later when the BH1-BH3 has already hardened significantly and the exchange of BH2 and BH3 is more violent. At last, the orbit of BH3 during the chaotic evolution stays very radial and experiences virtually no dynamical friction. The two latter reasons can be traced back to the slope of the central density distribution in the merger remnant since dynamical friction and orbital decay is more effective in systems with cuspy matter assembly \citep{Read_2006, Gualandris_2008, Just_2011, Antonini_2011, Ogiya_2020}. The absolute density at the center does not affect the triple formation strongly. Therefore, in a similar manner to the way that close SMBH pair formation is supported by the presence of dense cusps \citep{Tremmel_2018b, Khan_2018}, we expect that the formation of stable SMBH triples is enhanced if the galactic remnant has a central density cusp with sufficient steepness.

\section{Conclusion}
We identified three ROMULUS25 SMBH triple systems and recreated them for our high-resolution, gravitodynamical \N-body simulations. One system with a triple galaxy merger was investigated in more detail through three multiple initial conditions with varying central stellar density distributions. Our results show how in all cases the two heaviest SMBHs eventually form a hardening binary that coalesces within fractions of the Hubble time. The outcome for the lightest black hole depends sensitively on the initial condition. It can have a direct interaction with the heavier binary in the remnant center and may either form a stable hierarchical triple, if the central density distribution is sufficiently cuspy or be ejected from the stellar compound if the chaotic triple is very violent and takes place in an already flattened central matter distribution. In other instances, the third black hole has no direct interaction with the heavier binary and just merges with its remnant or stays on a stable wider galactic orbit insusceptible to dynamical friction. In any case, we find that the coalescence of the lightest black hole is delayed by the presence of the earlier heavier binary, as binary scouring reduces the magnitude of later dynamical friction and stellar hardening, and leads to a smaller mass ratio and hence longer gravitational wave coalescence. Three-body interactions between the heavier binary and lighter black hole may help to accelerate the binary's merger, but a KZL-like excitation of the inner eccentricity is usually suppressed by the relativistic periastron precession.

Our simulations show that additional exchanges and interactions alter the usual pattern of \citet{Begelman} when assessing the dynamical evolution of multiple SMBH systems. High-resolution \N-body simulations are a costly though necessary tool to address this problem, as exchanges and strong interactions are not really foreseeable when just looking at the dynamics on a wider scale from a cosmological simulation. Based on our results, we expect that observations of multiple AGN with large offsets could be in part the result of slingshot events when three or more SMBHs interact and a lighter SMBH is sent on a wider orbit where dynamical friction is low.

It remains open to investigate the interactions of a stable hierarchical triple with the field stars and how this affects the long-term hardening of the inner binary  and outer orbit evolution. So far, the usual questions about the dissipation of energy through stellar interactions and the influence of rotation or nuclear clusters have not been addressed for stable, hierarchical multiple SMBH systems in the literature. The large gap between the dimensionless hardening rates for the inner binary and outer orbits in the hierarchical triples of A.2 and A.3 (see Table \ref{tab:mergertimes}) suggests that the stellar hardening process for such a system could display very different characteristics than for the case of an SMBH binary. Additionally, studies including proper inclusion of three-body PN terms, velocity recoil, and spin evolution along the orbit are of vital interest \citep{Bonetti_2018, Mannerkoski_2021, Mannerkoski_2022} and seem necessary to understand the behavior of multiple SMBH systems as they occur in hierarchical galaxy formation.

\begin{acknowledgements}
We thank the referee for their insightful and helpful comments. 
The authors acknowledge support by the state of Baden-W\"urttemberg through bwHPC cluster BinAC at the Zentrum f\"ur Datenverarbeitung of the University of T\"ubingen. The authors gratefully acknowledge the Gauss Centre for Supercomputing e.V. (www.gauss-centre.eu) for funding this project by providing computing time on the GCS Supercomputer JUWELS \citep{JUWELS} at J\"ulich Supercomputing Centre (JSC).
We like to thank Yohai Meiron for his help with the \phiGRAPE\ setup on our computing clusters.
The work of PB was supported by the Volkswagen Foundation under the special stipend No.~9D154. 
PB acknowledge the support within the grant No.~AP14869395 of the Science Committee of the Ministry of Science and Higher Education of Kazakhstan (''Triune model of Galactic center dynamical evolution on cosmological time scale''). 
The work of PB was also supported under the special program of the NRF of Ukraine Leading and Young Scientists Research Support - ''Astrophysical Relativistic Galactic Objects (ARGO): life cycle of active nucleus'', No.~2020.02/0346.
The work of PB was supported under the special program ``Long-term program of support of the Ukrainian research teams at the PAS Polish Academy of Sciences carried out in collaboration with the U.S. National Academy of Sciences with the financial support of external partners''. 

\end{acknowledgements}

\bibliographystyle{aa}

\bibliography{references}  





\end{document}